\documentclass[b5paper,twoside]{jpconf}  
\usepackage{graphicx}

\newcommand{\labfig}[1]{\label{fig:#1}}

\newcommand{\fig}[1]{Figure~\ref{fig:#1}}

\begin{document}

\title[Morphology of galaxies]{Morphology of galaxies}  

\author[Y. Wadadekar]{Yogesh Wadadekar$^1$}

\address{$^1$National Centre for Radio Astrophysics, Tata Institute of Fundamental Research, Post Bag 3, Ganeshkhind, Pune 411007, India}

\ead{yogesh@ncra.tifr.res.in} 

\begin{abstract}

The study of the morphology of galaxies is important in order to
understand the formation and evolution of galaxies and their
sub-components as a function of luminosity, environment, and
star-formation and galaxy assembly over cosmic time. Disentangling the
many variables that affect galaxy evolution and morphology, requires
large galaxy samples and automated ways to measure morphology. The
advent of large digital sky surveys, with unprecedented depth and
resolution, coupled with sophisticated quantitative methods for
morphology measurement are providing new insights in this fast
evolving field of astronomical research.

\end{abstract}

The field of galaxy morphology has a long history in astronomy. It is
also a very wide and active field of research at the present time. A
recent review by Buta (\cite{but11}) is 174 pages long and cites about
350 papers. In this short review, due to paucity of space, only a
small part of this active field can be covered. I must admit that the
topics I cover are somewhat biased by my own research interests in
this area. For a more comprehensive and unbiased survey of the field,
the reader is referred to the excellent review by Buta
\cite{but11}. For a more pedagogical introduction, the classic text by
Binney and Merrifield (\cite{bin98}) is highly recommended.

\section{Galaxy morphology: a brief history}

It was realised nearly a century ago that galaxies were indeed
``island universes''; independent systems composed of a
gravitationally bound assemblage of stars, gas and dust. The study of
galaxy properties, began in earnest after this discovery. In the early
decades of the 20th century, it became clear that most bright galaxies
fell into two distinct categories - those with a smoothly declining
brightness distribution with no inflections, and no evidence for a
disc called ``ellipticals'' and the disc-dominated systems with spiral
arms punctuated with star-forming complexes, called ``spirals''. By
1936, when Hubble's book {\it Realm of the Nebulae} (\cite{hub36})
appeared, the study of galaxy morphology had become a well established
sub-field of optical astronomy. In this book, based on lectures he had
delivered at Yale University a year earlier, Hubble published the
Hubble sequence for galaxy classification (popularly known as the
``tuning fork diagram'', due to its resemblance to the shape of a
tuning fork). In Hubble's classification scheme (\fig{tuningfork}),
regular galaxies are divided into 3 broad classes - ellipticals,
lenticulars (S0) and spirals - based on their visual appearance on
photographic plates. A fourth class (added later, not seen in
\fig{tuningfork}) contains galaxies with an irregular appearance;
these were invariably forming stars at a rapid rate. Although a few
other other schemes of galaxy classification have been proposed in the
literature (e.g. \cite{mor58,van98}), it is the Hubble classification
(as revised and expanded by Sandage (\cite{san61}) and de Vaucouleurs
(\cite{dev59})) that is most widely used. According to Sandage
(\cite{san75}), one reason Hubble's view prevailed is that he did not
try and account for every superficial detail, but kept his classes
broad enough that the vast majority of galaxies could be sorted into
one of his proposed bins.

\begin{figure}
\begin{center}
\includegraphics[scale=0.2]{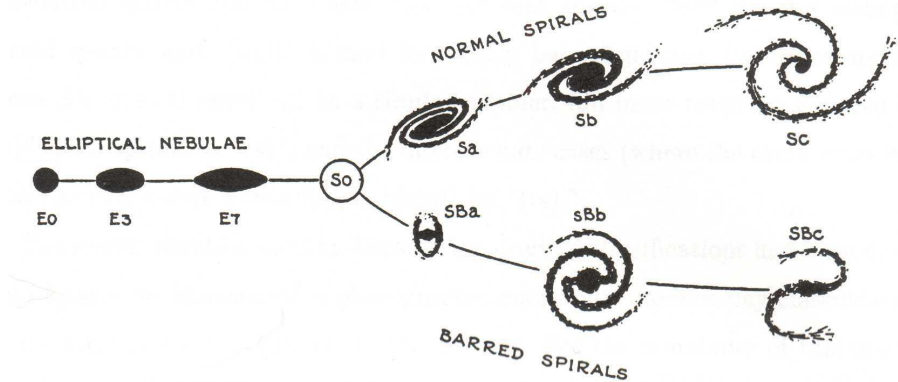}
\end{center}
\caption{Hubble's scheme for galaxy classification as it appeared in \cite{hub36}. Although more sophisticated versions of this scheme have been proposed by others, the basic ideas have survived for three quarters of a century.}
\labfig{tuningfork}
\end{figure} 

In recent years, the field of galaxy morphology has undergone a
renaissance for several reasons. These include:

\begin{enumerate}

\item Morphology is a fundamental property of galaxies. Any theory of
  galaxy formation and evolution has to explain the observed
  distribution of galaxies as a function of cosmic epoch and environment.

\item Galaxy morphology is strongly correlated with galactic star
  formation history. Galaxies where star formation ceased many
  gigayears ago usually have a different morphology from those where
  star formation continues at the present time (\fig{m51}). Galaxy
  morphology, therefore, is a zeroeth order tracer of star formation
  history.

\item Recent discoveries of new types of galaxies (\fig{greenpeas},
  \cite{car09}), and higher resolution views of nearby galaxies have 
  expanded the field as modern digital surveys and the Hubble Space Telescope
  have superseded the old photographic plates, that were in use for
  decades.

\item The explosion of data is accompanied by the development of
  quantitative techniques for automated measurement of galaxy
  morphology.

\item Visual classification of millions of galaxies has also been
  revolutionised by citizen science projects such as Galaxy
  Zoo\footnote{http://www.galaxyzoo.org}. Galaxy Zoo has transformed the field from the exclusive practice
  of a few experts to that of hundreds of thousands of enthusiastic Internet connected amateurs (without compromising on quality!).

\item The Hubble Space Telescope has enabled imaging studies of nearby
  galaxies at unprecedented resolutions (e.g. \fig{m51}) and deep
  surveys with the same telescope have extended morphological studies
  to $z=1$ and beyond (\fig{z3galaxies}).
\end{enumerate}

The present effort in the area is directed at obtaining an
understanding of how galaxy morphology is influenced by environmental
density, merger/interaction history, internal perturbations driven by
instabilities, gas accretion from other galaxies, nuclear activity,
internal secular evolution and star formation history (see
\cite{kor04} for a discussion of the interplay of all these
factors). Disentangling the effect of all these interconnected
influences on galaxy morphology is a complex exercise and is the
central problem of galaxy evolution.  Independent, yet synergistic
developments in 1. the development of theories of galaxy evolution
with predictions of observables such as galaxy morphology and
2. multiwavelength observations of large galaxy samples at a variety
of redshifts and in different environmental conditions (clusters,
groups, field) to test the predictions of the theories are enabling a
better understanding of galaxy evolution. It must be noted that,
increasingly, theories of galaxy evolution are developed as advanced
computer simulations that take into account all the relevant physics
of the gas, dust, stars and dark matter (e.g. \cite{bek11}). As
computing power has grown dramatically in the last two decades, the
simulations have become increasingly realistic.

\begin{figure}
\begin{center}
\includegraphics[scale=0.4]{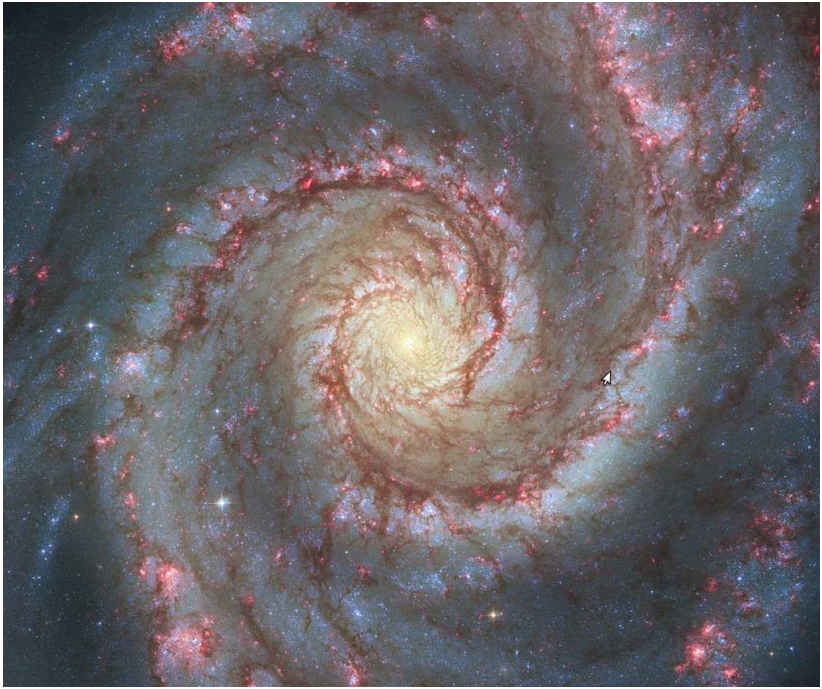}
\end{center}
\caption{Hubble Space Telescope image of the central regions of M51. High resolution imaging and the clear correlation between morphological features (spiral arms) and star formation complexes (red regions within the arms) make morphology a simple tracer of star formation. Image credit: S. Beckwith (STScI), Hubble Heritage Team, (STScI/AURA), ESA, NASA}
\labfig{m51}
\end{figure}

\begin{figure}
\begin{center}
\includegraphics[scale=0.5]{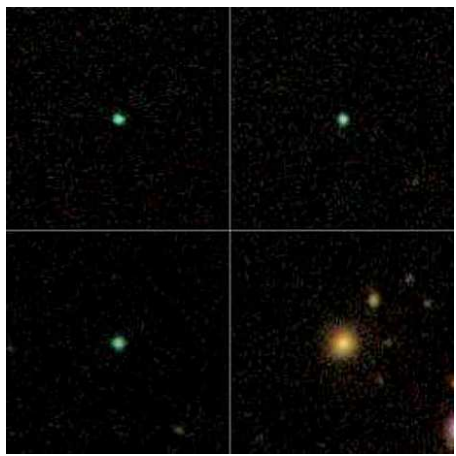}
\end{center}
\caption{A new class of round, green coloured galaxies labeled as {\it green peas} were discovered by volunteers in the Galaxy Zoo project. Peas are rare, no bigger than 5 kpc in radius, lie in lower density environments than normal galaxies, but may still have morphological characteristics driven by mergers. They are relatively low in mass and metallicity, and have a high specific star formation rate, yielding doubling times for their stellar mass of only hundreds of Myr (\cite{car09}).}
\labfig{greenpeas}
\end{figure}

\section{Quantitative morphology}

In the traditional method of classification, images of galaxies on
photographic plates (Kodak 103a-O and IIa-O were widely used) were
carefully examined by an expert, who then assigned a class to each
object. In blue sensitive plates, massive star clusters dominated by
early type stars stand out. At the same time dust absorption is severe
and provides a dramatic contrast to the star clusters. Galaxies with
spiral arms (where star formation and dust are both seen) are
therefore easy to classify while other types are not. There are
several other issues in working with photographic plates or their
digitised versions. These include:

\begin{itemize}

\item The visual classification process does not
  scale to large galaxy samples because of the limited availability of
  human experts. Large samples -- containing millions of galaxies --
  are the norm today with the availability of large area digital sky
  surveys such as the Sloan Digital Sky Survey \cite{aba09}.

\item Even experts tend to show a small subjective bias in their
  classification, which is difficult to quantify.

\item Faint, distant galaxies are very difficult to classify visually,
  since important guides to classification such as the presence of a
  disc or spiral arms may be hard to see visually, or may even be
  physically weak or absent in the earliest galaxies.
\end{itemize}

In such a situation, automated fitting and measurements of galaxy
morphology using digital images has rapidly become popular. The most
common approach involves extracting the structural parameters of a
galaxy by the separation of the observed light distribution into bulge
and disc components. The morphology can then be quantitatively
measured by computing the bulge to total luminosity ratio $B/T$.  The
ratio is close to 1 for disc-less ellipticals and systematically
decreases as one proceeds along the Hubble sequence, approaching a
value of close of zero for late-type spirals (Sd).  There is
considerable variation in the details of the decomposition techniques
proposed by various researchers. In recent years, methods that employ
2D fits to broad-band galaxy images have become popular
(e.g. \cite{wad99,pen10}). Most of these decomposition techniques
assume specific surface brightness distributions such as a generalised
de Vaucouleurs profile (\cite{dev48}) for the bulge and an exponential
distribution for the disc.

The bulge-disc decomposition essentially involves a numerical solution
to a signal-to-noise ratio ($S/N$) weighted minimisation problem.  The
technique involves iteratively building 2D image models that best fit
the observed galaxy images, with the quality of the fit quantified by
the $\chi^2$ value. Weights for the $\chi^2$ function are usually
computed using the S/N ratio at each pixel of the galaxy image.  The
model image needs to be convolved with the measured point spread
function (PSF) from the galaxy frame before the $\chi^2$ is computed
(\fig{pymorph}). The accuracy and reliability of the decomposition
procedure can be assessed using simulated galaxy images.  In addition
to permitting a fit to a bulge and disc light profile, most modern
codes allow one to fit for other structures such as a point source
(usually caused by the presence of an AGN at the galaxy centre) and a
bar. The most recent version of the widely used code {\it galfit} also
allows for fitting irregular, curved, logarithmic and power-law
spirals, ring, and truncated shapes (\cite{pen10}). Wrapper programs,
that enable fits to all galaxies in a specified image are useful to
obtain morphological properties for hundreds of galaxies, in one go
(\cite{vik10b}).

\begin{figure}
\begin{center}
\includegraphics[scale=0.5]{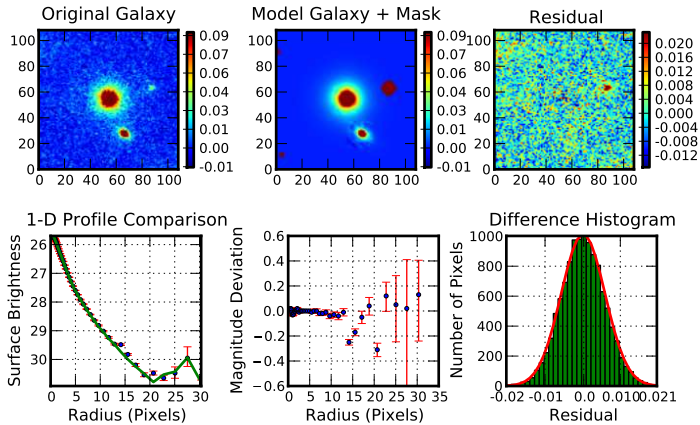}
\end{center}
\caption{Sophisticated graphical front-ends are now available to visualise the outputs of bulge-disc decomposition programs (\cite{vik10b}). Typically, a best fit signal-to-noise weighted analytic model of the 2D light profile of the galaxy is obtained. The model usually includes different galaxy components such as the bulge, disc, nuclear source, bars, spiral arms etc. One indicator of a good fit is when the residual (galaxy $-$ model) has a Gaussian (noise-like) distribution.}
\labfig{pymorph}
\end{figure} 

Once global parameters that describe the bulge and disc are available,
predicted correlations from theory can be tested against the
observations. I provide a couple of examples of how quantitative
morphology is improving our understanding of galaxy formation and
evolution.

\subsection{Evidence for luminosity dependent formation of lenticular bulges}

Lenticular (S0) galaxies straddle the space between ellipticals and
spirals in the Hubble tuning fork diagram (\fig{tuningfork}). It has
been clear for some time that bulges in ellipticals and late type
spirals are fundamentally different. Those in ellipticals seem to have
formed their stars rapidly at early epochs; while those in late-type
spirals have grown their bulges over time through internal evolution
processes such as secular evolution. Bulges of the elliptical kind
follow correlations such as the Kormendy relation and the Fundamental
Plane. Bulges in many spirals (called {\it pseudo bulges}), often show
correlated bulge and disk sizes indicating their formation through the
secular evolution mechanism (\cite{kor04}). In this context, it is
interesting to understand the formation process in the intermediate lenticular type. It has
been recently demonstrated that there seem to be two populations of
lenticular bulges differentiated by total luminosity of the
galaxies. Faint lenticulars show a positive correlation between bulge
and disc sizes, in line with predictions of secular formation
processes for the pseudo bulges of late-type disk galaxies. But
brighter lenticulars show an anti correlation, indicating that they
formed through a different mechanism (\cite{bar07}), most likely
involving major mergers. Galaxy environment also has an effect.  Faint
cluster lenticulars show systematic differences with respect to faint
field lenticulars. These differences support the idea that the bulge
and disc components fade after the galaxy falls into a cluster, while
simultaneously undergoing a transformation from spiral to lenticular
morphologies  (\cite{bar09}).

\begin{figure}
\begin{center}
\includegraphics[scale=0.4]{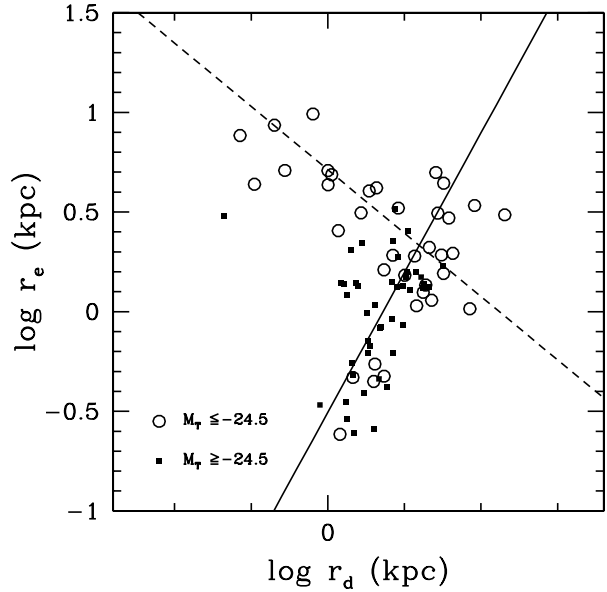}
\end{center}
\caption{Dependence of the bulge effective radius $r_e$ on the disc scale length $r_d$ for a sample of luminous and faint lenticulars. Dashed line is the best fit to the luminous lenticulars (circles) excluding five outliers, which shows an anti correlation. Solid line is the best fit to the less luminous lenticulars (squares) which show a positive correlation, indicating secular formation processes are active (\cite{bar07}).}
\labfig{lenticulars}
\end{figure} 

\subsection{Evolution of galaxy morphology in cores of clusters}

It has been known for some time that the fraction of early type
galaxies in the central regions of clusters has increased, as the
Universe evolved. With quantitative morphology measurements on HST
images of 379 galaxies in nine clusters spanning the redshift range 0.31
to 0.837, Vikram et al. (\cite{vik10a}) have recently measured the
fraction of bulge dominated galaxies, as a function of
redshift (\fig{bulgedominated}). They find a near monotonic decrease
with lookback time in the bulge-dominated fraction of galaxies; 40.0
$^{+2}_{-2}$ \% of galaxies at redshift $z = 0.837$ are bulge-like.
This increases to 55 $^{+3}_{-3}$ \% within $\sim$ 3.5 Gyr.

It must be noted that the trend above is weak and statistical in
nature; one needs to average over a large number of galaxies in a
large number of clusters over a wide range of redshift, to see a trend.
The detailed physics operating in each cluster, doubtless modifies the
morphological evolution of galaxies in that cluster. Nevertheless,
with a large, yet carefully selected galaxy sample, it is possible to
quantitatively measure changes which would be impossible to do
with a small sample.

\begin{figure}
\begin{center}
\includegraphics[scale=0.4]{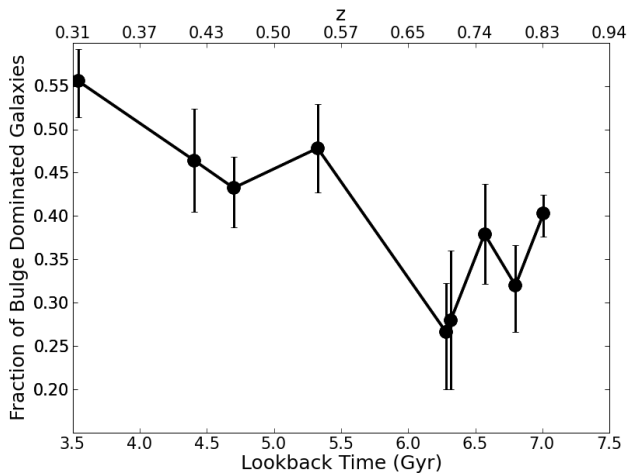}
\end{center}
\caption{Evolution of the fraction of bulge dominated galaxies in the cores of nine clusters with redshift in the range 0.31 to 0.837. The fraction of bulge dominated galaxies was lower when the Universe was younger (\cite{vik10a}).}
\labfig{bulgedominated}
\end{figure}

Work in both the above examples was enabled by the use of 2D bulge-disc
decomposition of galaxy images to measure quantitative parameters
describing the bulge and disc. The two examples quoted above are
merely representative of the work being done in this area. One has only to
glance through the large number of citations of \cite{pen02,pen10} to
get a feel for the enormous amount of research happening with quantitative
morphological measurements.

\section{Galaxy morphology at high redshifts}

Beyond $z \sim 1$, even with HST data, the parametric 2D bulge disk
decomposition technique does not work well. Besides the galaxies
appearing faint and small, the dropout selection technique frequently
used to find these distant galaxies, is biased towards highly star forming ones, which are
more likely to show disturbed morphologies (\fig{z3galaxies}).

To make classification possible at very high redshifts, several
non-parametric methods have been proposed and are widely used
(\cite{abr96,con03,lot04}. Non-parametric methods are not
computationally intensive compared to the parametric methods. However,
with non-parametric methods, it is not easy to convert measured
quantities to physically meaningful parameters such as bulge or disc
luminosity.

\begin{figure}
\begin{center}
\includegraphics[scale=0.5]{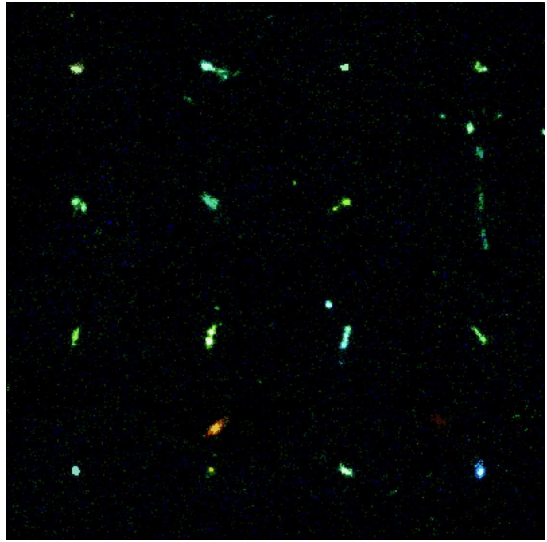}
\end{center}
\caption{At the centre of each of the 16 panels in the figure, is a star forming galaxy at $z \sim 3$ (\cite{wad06}) imaged with the GOODS survey on HST. At such early cosmic epochs, the well defined morphologies of galaxies in the nearby Universe are not seen, and an analytic decomposition of the light profile is unlikely to work well.}
\labfig{z3galaxies}
\end{figure}


\section*{References}
\begin{thebibliography}{4} 
\bibitem{aba09} Abazajian, K.~N., Adelman-McCarthy, J.~K., Ag{\"u}eros, M.~A., et al.\ 2009, \apjs, 182, 543
\bibitem{abr96} Abraham, R.~G., Tanvir, N.~R., Santiago, B.~X., et al.\ 1996, \mnras, 279, L47 
\bibitem{bar07} Barway, S., Kembhavi, A., Wadadekar, Y., Ravikumar, C.~D., \& Mayya, Y.~D.\ 2007, \apj, 661, L37
 \bibitem{bar09} Barway, S., Wadadekar, 
Y., Kembhavi, A.~K., \& Mayya, Y.~D.\ 2009, \mnras, 394, 1991 
\bibitem{bek11} Bekki, K., \& Couch, W.~J.\ 2011, \mnras, 415, 1783 
\bibitem{bin98} Binney, J., \& Merrifield, M.\ 1998, Galactic astronomy, Princeton: Princeton University Press
\bibitem{but11} Buta, R.~J.\ 2011, arXiv:1102.0550 
\bibitem{car09} Cardamone, C., Schawinski, K., Sarzi, M., et al.\ 2009, \mnras, 399, 1191 
\bibitem{con03} Conselice, C.~J.\ 2003, \apjs, 147, 1 
\bibitem{dev48} de Vaucouleurs, G.\ 1948, Annales d'Astrophysique, 11, 247
\bibitem{dev59} de Vaucouleurs, G.\ 1959, Handbuch der Physik, 53, 275 
\bibitem{hub36} Hubble, E.~P.\ 1936, Realm of the Nebulae, New Haven: Yale University Press, 1936
\bibitem{kor04} Kormendy, J., \& Kennicutt, R.~C., Jr.\ 2004, \araa, 42, 603
 \bibitem{lot04} Lotz, J.~M., Primack, J., \& Madau, P.\ 2004, \aj, 128, 163
 \bibitem{mor58} Morgan, W.~W.\ 1958, \pasp, 70, 364
\bibitem{pen02} Peng, C.~Y., Ho, L.~C., Impey, C.~D., \& Rix, H.-W.\ 2002, \aj, 124, 266 
\bibitem{pen10} Peng, C.~Y., Ho, L.~C., Impey, C.~D., \& Rix, H.-W.\ 2010, \aj, 139, 2097 
\bibitem{san61} Sandage, A.\ 1961, Hubble Atlas of Galaxies, Washington: Carnegie Institution
\bibitem{san75} Sandage, A. in Sandage, A., Sandage, M., \& Kristian, J. (eds.) \ 1975, Galaxies and the Universe, p. 1
\bibitem{van98} van den Bergh, S.\ 1998, Galaxy morphology and classification, Cambridge: Cambridge University Press
\bibitem{vik10a} Vikram, V., Wadadekar, Y., Kembhavi, A.~K., \& Vijayagovindan, G.~V.\ 2010, \mnras, 401, L39
\bibitem{vik10b} Vikram, V., Wadadekar, Y., Kembhavi, A.~K., \& Vijayagovindan, G.~V.\ 2010, \mnras, 409, 1379
\bibitem{wad99} Wadadekar, Y., Robbason, B., \& Kembhavi, A.\ 1999, \aj, 117, 1219 
\bibitem{wad06} Wadadekar, Y., Casertano, S., \& de Mello, D.\ 2006, \aj, 132, 1023
\end{thebibliography}
\end{document}